\documentclass[11pt]{article}
\usepackage[final]{acl}
\usepackage{times}
\usepackage{latexsym}
\usepackage[T1]{fontenc}
\usepackage{microtype}
\usepackage{hyperref}

\usepackage{graphicx}
\usepackage{subcaption}
\usepackage{booktabs}
\usepackage{tabularx}
\usepackage{ragged2e}
\usepackage{array}
\usepackage{amsmath}
\usepackage{amssymb}
\usepackage{mathtools}
\usepackage{amsthm}
\usepackage{multirow}
\usepackage{algorithm}
\usepackage{algorithmic}
\usepackage{tikz}
\usetikzlibrary{arrows.meta, positioning, shapes.geometric, fit, calc, decorations.pathreplacing}

\newcolumntype{Y}{>{\RaggedRight\arraybackslash}X}
\newcolumntype{C}[1]{>{\Centering\arraybackslash}p{#1}} 

\theoremstyle{plain}

\theoremstyle{definition}

\theoremstyle{remark}

\title{Speculative Refinement: A Hybrid Autoregressive Diffusion Decoding Strategy and Its Behavior Across Benchmarks}

\author{
  Aditi Gupta \\
  IIIT Hyderabad, India \\
  \texttt{\small{aditi.gu@research.iiit.ac.in}}
  \And
  Neel Mishra \\
  Microsoft \\
  \texttt{\small{mishraneel99@gmail.com}}
  \AND
  Kushagra Trivedi \\
  IIIT Hyderabad, India \\
  \texttt{\small{kushagra.trivedi@students.iiit.ac.in}}
  \And
  Pawan Kumar \\
  IIIT Hyderabad, India \\
  \texttt{\small{pawan.kumar@iiit.ac.in}}
}

\begin{document}
\maketitle

\begin{abstract}
How should we evaluate generation systems that combine autoregressive (AR) and diffusion decoding?
We study this question through \emph{Speculative Refinement} (SpecRef), a training-free hybrid method that warm-starts a masked diffusion language model from an AR draft using entropy-guided selective masking.
Evaluating SpecRef across six benchmarks (HumanEval, MBPP, GSM8K, BBH, ARC-Challenge, HellaSwag) with three distinct evaluation protocols (execution-based pass@1, exact-match, log-likelihood scoring), we surface several findings relevant beyond our specific system:
(1)~code benchmarks conflate \emph{structural discovery} with \emph{logical correctness}: providing a syntactic scaffold lifts accuracy from near zero to over 20\% without changing the model, indicating that much of the baseline failure is structural;
(2)~a \emph{refinement tension} phenomenon where multi-stage correction degrades already-correct tokens, exposing benchmark saturation ceilings invisible to single-model evaluation;
(3)~log-likelihood and generative evaluation produce different model rankings for the same model pair, suggesting they measure different capabilities;
(4)~standard Python post-processing silently breaks code evaluation for non-AR generators.
These observations apply to any multi-stage or non-autoregressive generation pipeline and point toward more diagnostic evaluation practices. Codes are \href{https://github.com/misterpawan/specref-eval.git}{here}.
\end{abstract}

\section{Introduction}

An 8-billion-parameter diffusion language model scores 0\% on HumanEval when given 16 denoising steps. Hand it the same problem with a syntactic scaffold from a 2.7B autoregressive drafter, and it scores over 20\% without any retraining. The model did not suddenly learn to code; the benchmark was testing whether it could discover Python's indentation rules from scratch, and that turned out to be the hard part. This is one of several findings that emerged when we tried to evaluate a hybrid generation system using standard benchmarks and discovered that the benchmarks themselves were not designed for this.

Diffusion language models (dLMs) have moved rapidly from research curiosity to a competitive generation paradigm. LLaDA~\citep{nie2025llada} demonstrated that masked diffusion scales to 8B parameters and matches LLaMA-3 on in-context learning. MDLM~\citep{sahoo2024mdlm} showed that simple masked diffusion with modern training recipes closes much of the perplexity gap with AR models. Dream~7B~\citep{ye2025dream} further narrowed this gap while retaining diffusion-native capabilities like parallel generation. On the commercial side, Inception Labs' Mercury~\citep{gulcehre2025mercury} achieved over 1,000 tokens per second on NVIDIA H100 GPUs while matching the quality of speed-optimized AR models, demonstrating that dLMs are viable at production scale. These models offer structural advantages that AR models lack: bidirectional context at every step, arbitrary position infilling, and the ability to correct multiple tokens in parallel. The question is no longer whether dLMs can work, but how to combine their strengths with the sequential planning ability of AR models.

Several recent works have begun exploring this direction. \citet{christopher2025specdiff} used diffusion as a drafter for speculative decoding. \citet{horvitz2024paraguide} conditioned diffusion paraphrasers on AR output. Warm-starting diffusion from corrupted inputs is well-studied in images~\citep{meng2021sdedit} and was recently formalized for language~\citep{scholz2025warmstarts}. These efforts suggest that hybrid AR-diffusion pipelines are a promising research direction, but they also raise a question that has received less attention: \emph{how should we evaluate them?}

The NLP community has built mature evaluation infrastructure~\citep{gehrmann2021gem,liang2023holistic}: execution-based code benchmarks~\citep{chen2021humaneval}, math reasoning suites~\citep{cobbe2021gsm8k}, logical deduction tasks~\citep{suzgun2023bbh}, and LLM-as-judge protocols~\citep{zheng2023judging}. But this infrastructure was built for left-to-right autoregressive (AR) decoding~\citep{vaswani2017attention}. Post-processing scripts assume tokens arrive in order. Execution sandboxes assume well-formed syntax. Log-likelihood scoring assumes a single forward pass. These assumptions break quietly when applied to dLMs~\citep{nie2025llada,austin2021d3pm,lou2024sedd}, speculative decoding pipelines~\citep{leviathan2023speculative,chen2023speculative}, and hybrid systems that combine both. Benchmark contamination~\citep{xu2024benchmarking} is already a known concern; we show that \emph{protocol mismatch} is equally serious and largely unexamined.

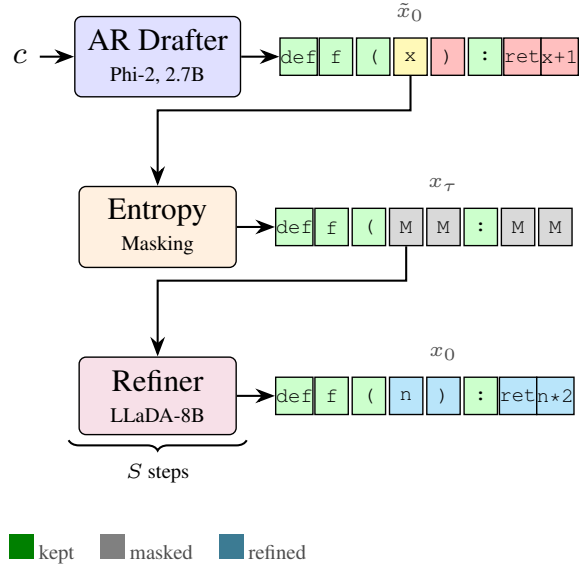
\begin{figure}[t]
\centering
\resizebox{\columnwidth}{!}{%
\begin{tikzpicture}[
    node distance=0.45cm,
    box/.style={draw, rounded corners=2pt, minimum height=0.7cm, minimum width=1.6cm, align=center, font=\footnotesize},
    tok/.style={draw, minimum width=0.34cm, minimum height=0.38cm, inner sep=0pt, font=\tiny\ttfamily},
    arrow/.style={-{Stealth[length=2mm]}, semithick},
    lbl/.style={font=\tiny, text=black!70},
]

\node[box, fill=blue!12] (drafter) {AR Drafter\\[-1pt]{\tiny Phi-2, 2.7B}};
\node[left=0.35cm of drafter, font=\footnotesize] (prompt) {$c$};
\draw[arrow] (prompt) -- (drafter);

\node[right=0.45cm of drafter] (d1) {};
\foreach \i/\c/\t in {1/green!20/def, 2/green!20/f, 3/green!20/(, 4/yellow!35/x, 5/red!25/), 6/green!20/:, 7/red!25/ret, 8/red!25/x+1} {
    \pgfmathsetmacro{\xoff}{(\i-1)*0.38}
    \node[tok, fill=\c] (t\i) at ([xshift=\xoff cm]d1) {\t};
}
\node[lbl, above=1pt of t4] {$\tilde{x}_0$};
\draw[arrow] (drafter) -- (t1);

\node[box, fill=orange!12, below=0.9cm of drafter] (entropy) {Entropy\\[-1pt]{\tiny Masking}};
\draw[arrow] (t4.south) -- ++(0,-0.35) -| (entropy.north);

\node[right=0.45cm of entropy] (m1) {};
\foreach \i/\c/\t in {1/green!20/def, 2/green!20/f, 3/green!20/(, 4/gray!30/{\tiny M}, 5/gray!30/{\tiny M}, 6/green!20/:, 7/gray!30/{\tiny M}, 8/gray!30/{\tiny M}} {
    \pgfmathsetmacro{\xoff}{(\i-1)*0.38}
    \node[tok, fill=\c] (m\i) at ([xshift=\xoff cm]m1) {\t};
}
\node[lbl, above=1pt of m5] {$x_\tau$};
\draw[arrow] (entropy) -- (m1);

\node[box, fill=purple!12, below=0.9cm of entropy] (refiner) {Refiner\\[-1pt]{\tiny LLaDA-8B}};
\draw[arrow] (m4.south) -- ++(0,-0.35) -| (refiner.north);

\node[right=0.45cm of refiner] (r1) {};
\foreach \i/\c/\t in {1/green!20/def, 2/green!20/f, 3/green!20/(, 4/cyan!25/n, 5/cyan!25/), 6/green!20/:, 7/cyan!25/ret, 8/cyan!25/n*2} {
    \pgfmathsetmacro{\xoff}{(\i-1)*0.38}
    \node[tok, fill=\c] (r\i) at ([xshift=\xoff cm]r1) {\t};
}
\node[lbl, above=1pt of r5] {$x_0$};
\draw[arrow] (refiner) -- (r1);

\draw[decorate, decoration={brace, amplitude=3pt, mirror}, semithick]
  ([xshift=-0.1cm, yshift=-0.05cm]refiner.south west) --
  ([xshift=0.1cm, yshift=-0.05cm]refiner.south east)
  node[midway, below=3pt, font=\tiny] (slabel) {$S$ steps};

\node[lbl, below=0.25cm of slabel] (leg) {%
  \textcolor{green!50!black}{\rule{0.25cm}{0.25cm}} kept \quad
  \textcolor{gray}{\rule{0.25cm}{0.25cm}} masked \quad
  \textcolor{cyan!50!black}{\rule{0.25cm}{0.25cm}} refined};

\end{tikzpicture}%
}
\caption{SpecRef pipeline. The AR drafter generates a candidate; per-token entropy identifies uncertain positions (yellow/red), which are replaced with \texttt{[MASK]}. The diffusion refiner denoises from this warm-started state in $S$ steps.}
\label{fig:pipeline}
\end{figure}

We expose these issues through \emph{Speculative Refinement} (SpecRef), a training-free hybrid method that uses a small AR model (Phi-2, 2.7B) to draft an initial answer, then selectively masks the least-confident tokens based on their Shannon entropy and lets a large diffusion model (LLaDA-8B~\citep{nie2025llada}) refine them (Figure~\ref{fig:pipeline}). The method works with any off-the-shelf AR drafter and requires no training. Evaluating it across six benchmarks with three evaluation protocols surfaced findings that go well beyond our specific system.

\paragraph{Contributions.} To the best of our knowledge, this is the first systematic study that pairs an AR language model as a drafter with a masked diffusion model (LLaDA) as a refiner, and evaluates the resulting hybrid across multiple benchmarks and evaluation protocols. Our main contributions are evaluation insights rather than the method itself:

\begin{itemize}
    \item \textbf{Structural discovery vs.\ logical correctness.} Code benchmarks (HumanEval~\citep{chen2021humaneval}, MBPP~\citep{austin2021mbpp}) conflate two capabilities: discovering syntactic structure (indentation, brackets, function signatures) and writing correct logic. Providing an AR scaffold separates these: accuracy jumps from near 0\% to 20\%+ at the same diffusion step budget (Table~\ref{tab:scaling_results}), indicating that a substantial portion of the standalone diffusion baseline's failure on code is structural rather than logical.
    \item \textbf{Refinement tension.} On benchmarks where the AR drafter is already accurate (BBH~\citep{suzgun2023bbh}: 82.8\%), multi-stage refinement can \emph{decrease} accuracy (to 80.4\%) because the refiner overwrites correct tokens. This ceiling effect is invisible when evaluating either model alone.
    \item \textbf{PPL vs.\ generative evaluation gap.} On multiple-choice tasks (ARC-Challenge, HellaSwag), log-likelihood scoring produces rankings that differ from those observed on generative benchmarks for the same model pair (Table~\ref{tab:ppl_results}), suggesting that PPL-mode and generative evaluation measure different capabilities in diffusion models.
    \item \textbf{Post-processing pitfalls.} Standard code post-processing (\texttt{.strip()}, \texttt{textwrap.dedent()}) silently destroys Python indentation for non-AR generators, lowering pass@1 scores. We document this and other practical pitfalls encountered when adapting AR-era evaluation to hybrid systems (Section~\ref{subsec:pitfalls}).
\end{itemize}

We also describe SpecRef itself (Section~\ref{sec:method}) since it is needed to interpret the evaluation findings, and report that it improves absolute accuracy by up to 25.8\% over standalone diffusion on code generation while cutting wall-clock latency by up to $2.43\times$.

\section{Background: Hybrid AR-Diffusion Generation}
\label{sec:method}

We describe the components of our hybrid system, which serves as the evaluation subject throughout the paper.

\subsection{AR drafter and uncertainty estimation}

Let $c$ denote a prompt and $x_0=(x_{0,1},\dots,x_{0,L})$ be a sequence of $L$ tokens from vocabulary $V$.
An AR drafter with parameters $\phi$ defines a left-to-right factored distribution:
\begin{equation}
q_\phi(x_0 \mid c) = \prod_{i=1}^{L} q_\phi(x_{0,i} \mid c, x_{0,<i}).
\end{equation}
At each position $i$, we record the Shannon entropy as a measure of uncertainty:
\begin{align}
u_i &\triangleq H(q_\phi(\cdot \mid c, x_{0,<i})) \notag\\
    &= -\!\sum_v q_\phi(v \mid \cdot) \log q_\phi(v \mid \cdot).
\end{align}
A high-entropy position means the AR model is unsure about that token, making it a natural candidate for diffusion refinement.
In our experiments we use \texttt{microsoft/phi-2} (2.7B parameters), but the method works with any AR model that outputs probability logits.

\subsection{Masked diffusion language modeling}

A masked diffusion LM such as LLaDA~\citep{nie2025llada} defines a forward process that progressively masks a clean token sequence toward a fully corrupted state using a token-level masking kernel:
\begin{equation}
q_t(x_{t,i} \mid x_{0,i}) = \begin{cases} x_{0,i} & \text{with prob.} \ (1-\alpha_t) \\ \texttt{[MASK]} & \text{with prob.} \ \alpha_t \end{cases},
\end{equation}
where $\alpha_t \in [0,1]$ is a monotone noise schedule with $\alpha_0 = 0$ (clean) and $\alpha_T = 1$ (fully masked).
The learned reverse process $p_\theta(x_{t-1}\mid x_t,c)$ is a bidirectional BERT-style Transformer (LLaDA-8B in our case) that attends to all positions at once and iteratively denoises from $t=T$ to $t=0$.
Unlike AR models, it can condition on ``future'' tokens at position $j > i$, and it can infill arbitrary subsets of tokens.

\subsection{SpecRef: warm-start with selective masking}

The main bottleneck when sampling from a fully masked prior is \emph{structure discovery}: at $t = T$, every token is \texttt{[MASK]}, so the model must recover code skeletons, reasoning chains, and answer formats from scratch. With a small step budget $S$, it often fails to produce syntactically valid output.

SpecRef warm-starts the diffusion model from an AR draft instead. It samples a draft $\tilde{x}_0 \sim q_\phi(\cdot\mid c)$, then applies the forward corruption kernel at time $\tau$:
\begin{equation}
x_\tau \sim q_\tau(\cdot \mid \tilde{x}_0),
\end{equation}
and runs the learned reverse process from $\tau$ to $0$:
\begin{equation}
x_{t-1} \sim p_\theta(\cdot \mid x_t, c)\quad \text{for}\ t=\tau,\tau{-}1,\dots,1.
\end{equation}
In practice, we use a subset of time indices (e.g., DDIM-style~\citep{song2021ddim}) to reduce the number of forward passes to $S \ll \tau$.

Rather than masking globally, SpecRef masks \emph{selectively}: only the top $k\%$ highest-entropy positions are replaced with \texttt{[MASK]}. Two heuristics extend the mask set $\mathcal{I}$:

\begin{enumerate}
    \item \textbf{Math-aware block expansion} ($e=2$): positions aligned to numeric or math-operator characters are identified via regex. All tokens within $\pm 2$ positions are added:
    $\mathcal{I} \leftarrow \mathcal{I} \cup \{i \mid |i - j| \le 2,\; j \in \mathcal{M}\}$,
    where $\mathcal{M}$ indexes math-character positions. This ensures arithmetic context is regenerated as a unit.
    \item \textbf{Tail truncation} ($M=150$): all tokens past index 150 are unconditionally masked to force the diffusion model to regenerate conclusions, where AR drift is worst.
\end{enumerate}

Token scores for the top-$k$ selection are:
\begin{equation}
\label{eq:token_score}
m_i = \mathrm{clip}\!\left(\frac{u_i}{\log |V|},\, 0,\, 1\right),
\end{equation}
where $u_i$ is mapped to refiner-token positions via character-level offset alignment (needed because the drafter and refiner use different tokenizers). We use $k=60\%$ throughout. Algorithm~\ref{alg:specref-token} gives the full procedure.

\begin{algorithm}[ht]
\caption{Statistical SpecRef (Top-\texorpdfstring{$k$}{k} Entropy Masking)}
\label{alg:specref-token}
\begin{algorithmic}[1]
\REQUIRE Prompt $c$; AR drafter $q_\phi$; refiner $p_\theta$; mask percentile $k$; math-expansion window $e$; tail truncation $M$.
\ENSURE Refined sample: $x_0$
\STATE Sample draft:  $\tilde{x}_0 \sim q_\phi(\cdot \mid c)$
\STATE Extract logit entropies:\\ $u_i = H(q_\phi(\cdot\mid c,\tilde{x}_{0,<i}))$
\STATE Compute the set of top $k\%$ tokens by entropy: $\mathcal{I}_{\text{top}}$
\STATE \textbf{Math Block Expansion}: Let $\mathcal{M}$ be indices of tokens aligned to numeric/math characters. Set $\mathcal{I} = \mathcal{I}_{\text{top}} \cup \{i \mid |i - j| \le e,\; j \in \mathcal{M}\}$
\STATE \textbf{Tail Truncation}: Set $\mathcal{I} = \mathcal{I} \cup \{i \mid i > M\}$

\STATE Form intermediate mask state $x_{\tau}$:
\[
x_{\tau,i} =
\begin{cases}
\texttt{[MASK]} & \text{if } i \in \mathcal{I} \\
\tilde{x}_{0,i} & \text{otherwise}
\end{cases}
\]
\FOR{refinement steps $S$}
  \STATE Sample proposal: $x'_{t-1} \sim p_\theta(\cdot \mid x_{t}, c)$
\ENDFOR
\STATE \textbf{return} $x_0$
\end{algorithmic}
\end{algorithm}

\subsection{Compute cost and the handoff point}

The total cost is $L\,C_{\text{AR}} + S\,C_{\text{D}}$, where $C_{\text{AR}}$ is the amortized cost of one AR token (negligible under vLLM serving) and $C_{\text{D}}$ is the cost of one diffusion forward pass over all $L$ positions. A standalone diffusion sampler with $N$ steps costs $N\,C_{\text{D}}$; SpecRef targets $S \ll N$.

The mask ratio $k\%$ (equivalently the warm-start time $\tau$) controls a tradeoff: $k \to 0$ preserves the entire draft; $k \to 100\%$ reduces to standard diffusion from scratch. We use $k=60\%$: enough tokens survive to carry global structure, while enough are masked for correction. On GSM8K (RTX 4090), this yields 6.79s per query vs.\ 16.49s for standalone LLaDA at $S=64$, a $2.43\times$ speedup.

\section{Evaluation Setup}
\label{sec:eval-setup}

\subsection{Models and infrastructure}

All experiments ran on the 8$\times$ NVIDIA V100, 16GB each. The refiner (\texttt{LLaDA-8B-Instruct}) is split across 2 GPUs via pipeline parallelism. The drafter (\texttt{Phi-2}, 2.7B) runs under vLLM with 4-bit quantization; its latency is under 1\% of total generation time.

\subsection{Evaluation protocols}
\label{subsec:protocols}

We deliberately chose benchmarks that require three different evaluation protocols, since the protocol itself turned out to be a significant variable.

\paragraph{Protocol A: Execution-based (pass@1).}
HumanEval~\citep{chen2021humaneval} and MBPP~\citep{austin2021mbpp}. Generated Python code is executed against ground-truth unit tests in a sandbox with a 10-second \texttt{SIGALRM} timeout (infinite loops are common in early diffusion steps). We report pass@1.

\paragraph{Protocol B: Exact match with parsing.}
GSM8K~\citep{cobbe2021gsm8k} and MATH (supplementary, evaluated on a 5,000-sample subset). Models are prompted to produce answers in \texttt{\textbackslash boxed\{...\}} format. We extract the final answer with regex and compute exact-match accuracy.

\paragraph{Protocol C: Log-likelihood scoring (PPL-mode).}
ARC-Challenge and HellaSwag. Each answer choice is formatted as \texttt{context + choice\_text}. The conditional log-likelihood is estimated via 8-round Monte Carlo masking (randomly masking a subset of continuation tokens per round, scoring under the model, and averaging), following the LLaDA evaluation protocol~\citep{nie2025llada}. The highest-scoring choice is selected. This protocol is diffusion-step-independent: no iterative denoising is performed, giving a $4\times$ speedup over generative evaluation.

\subsection{Evaluation pitfalls encountered}
\label{subsec:pitfalls}

Adapting standard AR-era evaluation to a hybrid system revealed several pitfalls that would affect any non-AR or multi-stage generator:

\begin{enumerate}
    \item \textbf{Post-processing breaks indentation.} Standard code post-processing (\texttt{.strip()}, \texttt{textwrap.dedent()}) destroys Python's semantic indentation. For diffusion outputs, which may contain irregular whitespace patterns unlike AR outputs, this lowered pass@1 scores significantly. We replaced these with stop-sequence truncation (\texttt{[\textbackslash{}ndef, \textbackslash{}nclass, \textbackslash{}nif \_\_name\_\_]}).
    \item \textbf{Infinite loops require sandboxing.} At low step counts ($S \le 16$), diffusion models frequently produce syntactically valid but non-terminating code. Without the \texttt{SIGALRM} timeout, evaluation hangs indefinitely, a problem AR models rarely cause.
    \item \textbf{Proxy metrics mislead for PPL-mode benchmarks.} During development, we tried string-matching the first 40 characters of the target as a proxy for PPL-mode benchmarks. The proxy and true metric disagreed substantially, producing misleading intermediate results.
    \item \textbf{Tokenizer mismatch.} The drafter and refiner use different tokenizers. Entropy scores must be mapped between tokenizations via character-level offset alignment; naive token-index transfer produces incorrect masks and unreliable evaluation.
\end{enumerate}

\section{Results}
\label{sec:results}

We organize results around evaluation findings rather than per-dataset performance. All numerical results are in Tables~\ref{tab:scaling_results}--\ref{tab:latency}; we do not repeat numbers already in the tables except when interpreting them.

\subsection{What code benchmarks actually test: structural discovery}

\begin{table*}[t]
\centering
\caption{Step-wise scaling on generative benchmarks. SpecRef sets a quality floor at low step counts ($S=8, 16$) where standalone diffusion fails. All results use \texttt{microsoft/phi-2} as drafter and \texttt{GSAI-ML/LLaDA-8B-Instruct} as refiner.}
\label{tab:scaling_results}
\resizebox{\textwidth}{!}{
\begin{tabular}{l c cccc cccc}
\toprule
\multirow{2}{*}{\textbf{Dataset}} & \multirow{2}{*}{\textbf{Phi-2 (Draft)}} & \multicolumn{4}{c}{\textbf{LLaDA Baseline}} & \multicolumn{4}{c}{\textbf{SpecRef (Ours)}}\\
\cmidrule(lr){3-6} \cmidrule(lr){7-10}
 & & \textbf{S=8} & \textbf{S=16} & \textbf{S=32} & \textbf{S=64} & \textbf{S=8} & \textbf{S=16} & \textbf{S=32} & \textbf{S=64} \\
\midrule
GSM8K & 47.38\% & 8.72\% & 12.43\% & 22.14\% & 44.81\% & 39.50\% & 45.26\% & 50.95\% & \textbf{54.28\%} \\
HumanEval & 44.51\% & 0.00\% & 0.00\% & 1.74\% & 8.70\% & 20.12\% & 20.73\% & 20.73\% & \textbf{25.61\%} \\
MBPP & 43.00\% & 0.20\% & 0.60\% & 2.00\% & 7.20\% & 20.80\% & 22.80\% & 26.00\% & \textbf{33.00\%} \\
MATH* & 5.22\% & 10.76\% & 12.72\% & 16.12\% & \textbf{22.32\%} & 14.16\% & 17.70\% & 20.44\% & 21.38\% \\
BBH (Bool) & 82.80\% & 54.80\% & 53.60\% & 58.80\% & 66.00\% & 71.20\% & 76.80\% & 80.40\% & \textbf{80.40\%} \\
\bottomrule
\end{tabular}
}
\begin{flushleft}
\footnotesize{* Note: The MATH evaluation was performed on a 5,000-sample representative subset of the final test suite.}
\end{flushleft}
\end{table*}

The most striking pattern in Table~\ref{tab:scaling_results} is on code benchmarks. Standalone LLaDA produces 0\% valid code on HumanEval at $S \le 16$ and only 8.70\% at $S=64$. But when warm-started from an AR draft, SpecRef reaches 20.12\% at just 8 steps, without changing the diffusion model at all.

This gap tells us something about what HumanEval and MBPP actually measure. This suggests the diffusion model at 8B parameters is not entirely incapable of code logic; rather, it cannot discover the right syntactic structure (indentation, brackets, function signatures) from a fully masked starting point in a few steps. Once that structure is provided, accuracy jumps immediately. These benchmarks appear to conflate two distinct capabilities: \emph{structural discovery} and \emph{logical correctness}. For diffusion models at low step counts, the structural component accounts for a large share of the failure.

A near-zero score on HumanEval does not mean a diffusion model ``cannot code.'' It may mean the model cannot discover Python's formatting conventions quickly enough. Separating structural and logical evaluation (e.g., providing scaffolds and measuring only logic fill-in) would give more diagnostic information.

On MBPP the pattern is even clearer: SpecRef's absolute gain expands to 25.80\% at $S=64$ (33.00\% vs.\ 7.20\%). On GSM8K, SpecRef at $S=64$ reaches 54.28\%, outperforming both the standalone drafter (47.38\%) and refiner (44.81\%), suggesting complementary strengths on math reasoning.

\subsection{Refinement tension: when correction hurts}

A second finding concerns what happens when the AR drafter is already accurate. On BBH Boolean Expressions~\citep{suzgun2023bbh}, the drafter scores 82.80\%. SpecRef improves over the standalone diffusion baseline at every step count but plateaus at 80.40\% for $S \ge 32$, below the drafter itself. The same pattern appears on MATH, where the baseline edges ahead of SpecRef at $S=64$ (22.32\% vs.\ 21.38\%).

We call this \emph{refinement tension}: the entropy-based mask sometimes selects tokens that the drafter already had correct, and the refiner replaces them with worse alternatives. The drafter's entropy is a noisy signal of actual error; on saturated tasks, high entropy does not always mean the token is wrong.

This phenomenon is invisible when evaluating either model in isolation. It only appears in the pipeline, and it reveals a benchmark saturation ceiling: once a component model is near-ceiling, aggregate pipeline scores can \emph{decrease}. For evaluation of any multi-stage system (not just SpecRef), this argues for reporting component-level scores alongside pipeline scores. It also suggests that adaptive, per-instance mask rate selection (lowering $k$ when the drafter is already confident) could mitigate the effect, though we leave this to future work.

\subsection{The log-likelihood vs.\ generative evaluation disconnect}

\begin{table}[ht]
\centering
\caption{Likelihood (PPL) performance. SpecRef in PPL mode matches the baseline because both use log-likelihood scoring on the full sequence.}
\label{tab:ppl_results}
\small
\begin{tabular}{l p{1.3cm} p{1.3cm} p{1.1cm}}
\toprule
\textbf{Dataset} & \textbf{Phi-2 (Draft)} & \textbf{Baseline (S=64)} & \textbf{SpecRef} \\
\midrule
ARC-Chall. & 58.10\% & 51.45\% & 51.50\% \\
HellaSwag & 54.94\% & 45.62\% & 45.62\% \\
\bottomrule
\end{tabular}
\end{table}

Table~\ref{tab:ppl_results} shows PPL-mode evaluation on ARC-Challenge and HellaSwag. Two observations are worth noting.

First, PPL-mode is diffusion-step-independent. The same score emerges regardless of step count, because log-likelihood is computed in a single masked forward pass, not through iterative denoising. This makes PPL-mode evaluation $4\times$ cheaper on our cluster, but it also means PPL-mode and generative evaluation measure fundamentally different things. Whether a model can \emph{score} a correct answer highly is not the same as whether it can \emph{generate} that answer.

Second, the AR drafter outperforms LLaDA on HellaSwag in PPL mode (54.94\% vs.\ 45.62\%), yet on generative benchmarks (Table~\ref{tab:scaling_results}) the relationship between the two models is more complex and step-dependent. While these are different tasks, the pattern suggests that PPL-mode rankings do not straightforwardly transfer to generative settings. For diffusion models specifically, PPL-mode may overstate or understate generation ability depending on the task.

\subsection{Latency-quality tradeoffs}

\begin{table}[ht]
\centering
\caption{Latency (seconds per query) on GSM8K, NVIDIA RTX 4090. SpecRef times include both the AR drafting pass (vLLM) and the diffusion steps. Because fewer diffusion steps are needed to match baseline quality, total wall-clock time drops.}
\label{tab:latency}
\small
\begin{tabular}{l p{1.7cm} p{1.5cm} p{1.2cm}}
\toprule
\textbf{Steps ($S$)} & \textbf{LLaDA Baseline (s)} & \textbf{SpecRef (s)} & \textbf{Speedup Factor} \\
\midrule
16 & 4.12 & 5.18 & $0.80\times$ \\
32 & 8.24 & 5.95 & $1.38\times$ \\
64 & 16.49 & 6.79 & \textbf{2.43$\times$} \\
\bottomrule
\end{tabular}
\end{table}

Table~\ref{tab:latency} shows that SpecRef $S=64$ is both more accurate and $2.43\times$ faster than standalone LLaDA at the same step count. At $S=16$SpecRef, it is slower (the drafting pass dominates when diffusion steps are few) but far more accurate.

For evaluation methodology, the lesson is that \textbf{reporting only accuracy is insufficient} for comparing generation paradigms with different cost structures. Step count, wall-clock time, and FLOPs all affect the comparison. Table~\ref{tab:scaling_results} shows accuracy swings of over 30 percentage points across step counts for the same model; a single-step-count comparison could be arbitrarily misleading.

\section{Discussion: Lessons for Evaluating Hybrid Generation}

\paragraph{1. Report evaluation protocol details exhaustively.}
Post-processing choices alone can change code evaluation scores by large margins (Section~\ref{subsec:pitfalls}). Sandbox timeouts, stop-sequence selection, and tokenizer alignment details are not minor implementation notes; they are evaluation-critical parameters. We recommend treating them with the same care as hyperparameters.

\paragraph{2. Benchmark scores conflate multiple capabilities.}
Code benchmarks test structural discovery and logical correctness simultaneously. For AR models, this distinction rarely matters (structure comes for free from left-to-right generation), but for diffusion and hybrid models, it is the dominant factor. Providing syntactic scaffolds as a controlled experiment could separate these components and give more diagnostic evaluations.

\paragraph{3. Multi-stage systems need component-level reporting.}
Aggregate pipeline scores can mask component-level regression (the refinement tension effect). We recommend reporting drafter-only, refiner-only, and pipeline scores together, along with an analysis of where the pipeline helps and where it hurts.

\paragraph{4. PPL-mode and generative evaluation are not interchangeable.}
They measure different aspects of model capability and can rank the same models differently. Papers that use only one mode should be explicit about this limitation.

\paragraph{5. Step count and compute budget must be reported.}
Diffusion model accuracy varies by 30+ pp across step counts (Table~\ref{tab:scaling_results}). A score without a step count is incomplete; accuracy-vs-compute curves should replace single numbers.

\section{Related Work}

GEM~\citep{gehrmann2021gem} and HELM~\citep{liang2023holistic} established shared evaluation infrastructure; LLM-as-judge methods~\citep{zheng2023judging} added scalable evaluation beyond reference matching; and benchmark contamination~\citep{xu2024benchmarking} has motivated living benchmarks. \citet{deschenaux2024promises} noted evaluation challenges for discrete diffusion but did not systematically document them, our work fills this gap. Continuous-space diffusion for text was introduced by \citet{li2022diffusionlm}; discrete extensions~\citep{austin2021d3pm,ho2020ddpm,song2021scorebased} led to SEDD~\citep{lou2024sedd} and MDLM~\citep{sahoo2024mdlm}. Speculative decoding~\citep{leviathan2023speculative,chen2023speculative} accelerates AR generation via draft-then-verify; SpecRef inverts this pattern, adapting warm-start diffusion~\citep{meng2021sdedit,scholz2025warmstarts} to discrete masked diffusion with a confidence-aware corruption policy.

\section{Conclusion}

We used SpecRef as a lens to examine how standard benchmarks behave when generation is no longer purely autoregressive, surfacing four findings: code benchmarks conflate structural discovery with logical correctness; multi-stage pipelines can degrade accuracy through refinement tension; log-likelihood and generative evaluation are not interchangeable; and standard post-processing silently breaks evaluation for non-AR generators. We hope these observations contribute to more diagnostic evaluation practices as generation architectures continue to diversify.

\section*{Limitations}
\label{sec:limitations}

Our evaluation uses a single drafter-refiner pair (Phi-2 2.7B + LLaDA 8B); quantitative magnitudes are specific to this combination, though the qualitative findings likely generalize. Phi-2's 2,048-token context window prevented evaluation on long-prompt benchmarks (e.g., GPQA). The fixed mask rate $k=60\%$ was not tuned per task; adaptive masking may reduce refinement tension.

\section*{Acknowledgement}
\label{sec:ack}
This work was carried out at IIIT, Hyderabad. We thank the insitute for logistics and compute resources. For GPUs rentals, we used vast.ai, we acknowledge the seamless access. This work was funded by Qualcomm faculty award at IIIT, Hyderabad for funding the first author. We thank Qualcomm for such unrestricted grant to support research.

\section*{Ethics Statement}
\label{sec:ethics}

SpecRef is a decoding strategy; it does not change training data or model capabilities.
It does, however, reduce the cost of generating text, which can amplify both beneficial and harmful uses.
Risks include:
\begin{itemize}
\item \textbf{Scaling misuse:} faster generation can facilitate spam, disinformation, or automated harassment.
\item \textbf{Bias propagation:} refinement can preserve biases from either model, and selective masking may ` "lock in" problematic phrasing from the draft.
\item \textbf{Over-trust in refinement:} users may assume ``refined" outputs are more factual than they are.
\end{itemize}
Mitigations include maintaining standard safety filters, auditing outputs for bias/toxicity, and using evaluation prompts that stress factuality and sensitive domains.

\bibliography{refs}

\end{document}